\input harvmac.tex
\vskip 1.5in
\Title{\vbox{\baselineskip12pt
\hbox to \hsize{\hfill}
\hbox to \hsize{\hfill}}}
{\vbox{\centerline{Conformal Moduli 
and $b-c$ Pictures for NSR Strings}
\vskip 0.2in
{\vbox{\centerline{}}}}}
\centerline{Dimitri Polyakov\footnote{$^\dagger$}
{dp02@aub.edu.lb}}
\medskip
\centerline{\it Center for Advanced Mathematical Studies\footnote
{$^\dagger$}{associate member}}
\centerline{\it and  Department of Physics }
\centerline{\it  American University of Beirut}
\centerline{\it Beirut, Lebanon}
\vskip .4in
\centerline {\bf Abstract}
We explore the geometry of superconformal moduli
of the NSR superstring theory in order to construct
the consistent sigma-model for NSR strings, free of picture-changing
complications.The sigma-model generating functional
 is constructed by the integration over the bosonic and fermionic
moduli, corresponding to insertions of the vertex operators
in scattering amplitudes.While
the integration over the supermoduli leads to the standard picture-changing 
insertions, the integration over the bosonic moduli results
in the appearance of  
picture-changing operators for the $b-c$ fermionic ghosts
with the ghost number $-1$.
Important example of the $b-c$ ghost pictures involves
the vertex operators in integrated and unintegrated forms.
We obtain the BRST-invariant expressions for the 
$b-c$ picture-changing operators for open and closed strings
and study some of their properties. 
We also show that the superconformal moduli 
spaces of the NSR theory contain the global singularities, leading
to the phenomenon of ghost-matter mixing and the appearance of
non-perturbative  D-brane  creation operators.
\Date{January 2004}

\vfill\eject
\lref\self{D.Polyakov, Phys.Rev.D65:084041,2002, hep-th/0111227}
\lref\verlinde{H Verlinde, Phys.Lett. B192:95 (1987)}
\lref\myself{D.Polyakov, in progress}
\lref\dhoker{E D'Hoker, D.H. Phong, 
hep-th/0110283, Nucl. Phys. B636:3-60 (202)}
\lref\dhokerp{E. D'Hoker, D.H. Phong, hep-th/0111016,
Nucl. Phys. B636:61 (2002)}
\lref\kp{I.I. Kogan, D.Polyakov, hep-th/0208036, Int. J. Mod. Physics
A18:1827,2003}
\lref\fts{E. Fradkin, A.Tseytlin, Nucl. Phys B261:1-27 (1985)}
\lref\fms{D. Friedan, E. Martinec, S. Shenker, Nucl. Phys. B271:93, 1986}
\lref\ampf{S.Gubser,I.Klebanov, A.M.Polyakov,
{\bf Phys.Lett.B428:105-114}}
\lref\malda{J.Maldacena, Adv.Theor.Math.Phys.2 (1998)
231-252, hep-th/9711200}
\lref\pols{J.Polchinski, L.Susskind, N.Toumbas, hep-th/9903228}
\lref\kn{D. Knizhnik, Usp. Fiz. Nauk, 159:401-453 (1989)}
\lref\koba{Z.Koba, H.Nielsen, Nucl. Phys B12:517-536 (1969)}
\lref\witten{E.Witten, Adv. Theor. Math. Phys.2:253 (1998)}
\lref\pol{J.Polchinski, Phys.Rebv. Lett.75:4724(1995)} 
\lref\nohta{N.Ohta, Phys. Rev. D33:1681 (1986)}
\lref\nnohta{N. Ohta, Phys. Lett. B179:347 (1986)}

\centerline{\bf 1. Introduction}

Despite the fact
that the NSR (Neveu-Schwarz-Ramond) superstring theory
appears to be the best explored, elegant and relatively simple
model describing the superstring dynamics, it still suffers a number
of serious complications. One, and perhaps the most important of them is
related to the problem of picture-changing and
the  difficulties in constructing the generating functional
for Ramond and Ramond-Ramond states, because of the picture-changing 
ambiguity ~{\fms}. Since the
Ramond vertex operators have fractional ghost-numbers,
one always  has to consider their combinations at different
ghost pictures in order to calculate their scattering amplitudes
(except for the 4-point function where all the operators
can be taken at the picture $-1/2$). As a result, constructing the
generating functional for the Ramond scattering amplitudes becomes
a confusing question. Actually,
the problem of the picture-changing is not limited 
to the Ramond sector.Due to the ghost number anomaly cancellation condition,
vertex operators of any correlation function in NSR string theory must 
have total superconformal ghost number $-2$
and total $b-c$ ghost number 3. This means that we cannot limit ourselves
to just integrated and picture  $0$ vertices in the generating functional
of the NSR model, since both integrated and unintegrated vertex operators,
at various superconformal pictures must be involved in the sigma-model
action. Therefore, due to this picture
ambiguity, we face the problem building the generating functional
for the consistent perturbation theory for NSR strings.
In this paper we show how this problem can be resolved if one accurately takes
into account the dependence of the superstring correlators on the
superconformal moduli and explores the geometry of supermoduli spaces, related
to the vertex operator insertions on the sphere or the disc.
The paper is organized as follows.
In the first section, we construct the 
basis for the
supermoduli related to the vertex operator insertions
and study the geometry 
of the related supermoduli spaces.
In the second section we perform the integration over the supermoduli
leading to the appearance of the bosonic and $b-c$ ghost picture changing.
Integrated and unintegrated vertex operators are shown to be the
particular examples of different $b-c$ pictures.
The general BRST-invariant expression for the $b-c$ picture-changing is 
derived.
We show that if one takes all the vertex operators of the sigma-model
 unintegrated and
 at the picture $-1$, the supermoduli integration insures the correct
$b-c$ and $\beta-\gamma$ ghost number balance. 
In the third section we study the 
global singularities of the supermoduli spaces, leading to the
phenomenon of the ghost-matter mixing ~{\self, \kp}
 and the appearance of the non-perturbative  vertices carrying
the RR-charges, which can be interpreted as the D-brane creation operators.
We conclude by introducing
the general expression for the NSR sigma-model, free of picture-changing 
ambiguity,by using the relevant superconformal moduli structures.
In the concluding section
 we discuss some possible implications of our results.
 
\centerline{\bf NSR sigma-model and moduli integration}

Consider the string scattering amplitudes in the
NSR formalism.To be certain, let us  first consider the closed string case,
the open strings
can be treated similarly.
The scattering amplitude on a sphere for N vertex operators
in the  NSR superstring theory is given by:
\eqn\grav{\eqalign{
<V_1(z_1,{\bar{z_1}})...V_N(z_N,{\bar{z_N}})>
=\int{\prod_{i=1}^{M(N)}}dm_id{\bar{m_i}}\int{\prod_{a=1}^{P(N)}}d\theta_a
d{\bar{\theta_a}}\int{DX}D{\psi}D{\bar\psi}D{\lbrack{ghosts}\rbrack}
\cr
e^{-S_{NSR}+m_i<\xi^i|T_{m}+T_{gh}>+{\bar{m_i}}<{\bar{\xi^i}}|{\bar{T}}_m
+{\bar{T}}_{gh}>+
\theta_a<\chi^a|G_m+G_{gh}>+{\bar{\theta}_a}<{\bar{\chi}}_a|{\bar{G}}_m+
{\bar{G}}_{gh}>}\cr
{\prod_{a=1}^{M(N)}}\delta(<\chi^a|\beta>)\delta(<{\bar{\chi}}^a|{\bar\beta}>)
{\prod_{i=1}^{P(N)}}
<\xi^i|b><{\bar\xi}^i|{\bar{b}}>V_1(z_1,{\bar{z}}_1)...V_N(z_N,{\bar{z}}_N)}}
Here
$z_1,...,z_N$ are the points of the vertex operator insertions
on the sphere and
\eqn\grav{\eqalign{S_{NSR}\sim\int{d^2z}\lbrace
\partial{X_m}\bar\partial{X^m}+\psi_m\bar\partial\psi^m
+\bar\psi_m\partial\bar\psi^m+b\bar\partial{c}+\bar{b}\partial{\bar{c}}
+\beta\bar\partial\gamma+\bar\beta\partial\bar\gamma\rbrace\cr
m=0,...,9
}}
is the NSR superstring action in the superconformal gauge.
Next, $(m_i,\theta_a)$ are the holomorphic even and odd coordinates in the
moduli superspace and $(\xi^i,\chi^a)$ are their dual super Beltrami
differentials (similarly for $({\bar{m_i}},{\bar\theta_a})$ and
$({\bar\xi^i},{\bar\chi^a})$
The $<...|...>$ symbol stands for the scalar
product in the Hilbert space and the delta-functions
$\delta(<\chi^a|\beta>)$ and $\delta(<\xi^i|b>)=<\xi^i|b>$
are needed to insure that the basis in the moduli space
is normal to variations along the superconformal gauge slices
(similarly for the antiholomorphic counterparts).
The dimensions of the moduli and supermoduli spaces are related to
the number N
of the vertex operators present and are different for
 the NS and Ramond sectors (see the discussion below).
The standard bosonization relations for the $b,c,\beta$ and $\gamma$
ghosts are given by ~{\fms}
\eqn\grav{\eqalign{c(z)=e^{\sigma}(z);
b(z)=e^{-\sigma}(z)\cr
\gamma(z)=e^{\phi-\chi}(z);\beta(z)=e^{\chi-\phi}\partial\chi(z)\cr
<\sigma(z)\sigma(w)>=<\chi(z)\chi(w)>=-<\phi(z)\phi(w)>=log(z-w)}}
The explanation is needed for the formula (1).
In NSR superstring theory, while the partition function
on the sphere does not have of course any modular dependence,
 the integration over
the moduli and the supermoduli does appear for the sphere scattering
amplitudes, for all the N-point correlators with
$N{\geq}3$.
This is related to the important fact that in string theory the
BRST and the local gauge invariances are not isomorphic to each other
{\nohta,\nnohta}.
 Namely,
the physical vertex operators
while being BRST-invariant, are not necessarily invariant
under the local supersymmetry. Moreover, while the integrated
vertices are invariant under the local reparametrizations,
the unintegrated ones are not. E.g. the infinitezimal
conformal transformation of an unintegrated
photon at zero momentum gives:
\eqn\grav{\eqalign{\delta_\epsilon{V_{ph}(z)}=
\oint{{dw}\over{2i\pi}}\epsilon(w)T(w)({c\partial{X^m}}+\gamma\psi^m)(z)
\cr=\oint{{dw}\over{2i\pi}}\epsilon(w)\lbrace
{1\over{z-w}}\partial(c\partial{X^m}+\gamma\psi^m)(z)
+O((z-w)^0)\rbrace=\epsilon(z)\partial(c\partial{X^m}+\gamma\psi^m)(z)}}
More generally, under the infinitezimal conformal transformations
$z\rightarrow{z+\epsilon(z)}$
any dimension 0 primary field transforms as
$$\Phi(z)\rightarrow\Phi(z+\epsilon(z))=\Phi(z)+\epsilon\partial\Phi(z)$$
Similarly, applying the local supersymmetry generator
 $\oint{{dw}\over{2i\pi}}\epsilon(w)G(w)$ to
a photon at picture $-1$: $V_{ph}=ce^{-\phi}\psi^m$
with G being a full matter$+$ghost supercurrent
and $\epsilon$ now a fermionic parameter,
it is easy to calculate
$$\delta_\epsilon{V_{ph}}=\epsilon(z)(-{1\over2}c{e^{-\phi}}\partial{X^m}
+c\partial{c}e^\chi{\partial\chi}e^{-2\phi}\psi^m)\neq{0}.$$
Note, however, that the photon vertex operator at picture $0$
is supersymmetric.
The problem is that, due to the $b-c$ and $\beta-\gamma$
ghost number anomalies on the sphere we cannot limit ourself
to just integrated vertices or those at the picture 0
but must always consider a combination of pictures in the amplitudes.
This combination, however, is not arbitrary
but must follow from the consistent
superstring perturbation theory, or the expansion in the appropriate
sigma-model terms. These terms are typically of the form
$\lambda{V}$ where V is a vertex operator of some physical state
and $\lambda$ is the space-time field corresponding to this vertex
operator.This term must be added to the original action as a string emits
the $\lambda$-field. Then the string partition function must be
expanded in $\lambda$ leading to the perturbation theory series
and the effective action for $\lambda$, upon computing the  correlators
~{\fts}.
But the important question 
is - which vertex operator should we take? Should it be integrated or
unintegrated, which superconformal ghost picture should be chosen?
If we choose the integrated and picture 0 vertices
as the sigma-model terms, the
 local superconformal symmetry is preserved, but we face the problem of
the ghost number anomaly cancellation. For this reason these
vertex operators are not suitable as the sigma-model terms.
If, on the other hand, we choose unintegrated vertices at the picture -1,
then one faces the question of the gauge non-invariance -
though these operators are BRST-invariant, they are not invariant
under local superconformal gauge transformations.
Therefore, in order to insure the gauge invariance of the
scattering amplitudes, it is necessary to impose some restrictions
on the parameter $\epsilon(z)$ of the superconformal transformations;
namely, it must vanish at all the insertion points of the vertex operators.
With such a restriction on the gauge parameter (effectively
reducing the superconformal gauge group) the scattering amplitudes
will be gauge invariant.
Below we will prove that the consistent superstring perturbation theory
must be defined as follows:

1) All the vertex operators must be taken unintegrated
and at the picture $-1$.

2)The restrictions on the gauge parameter:
\eqn\lowen{\epsilon(z_k)=0;k=1,...N}
must be imposed at the insertion points of the vertex operators
(where $\epsilon$ is either a bosonic infinitezimal reparametrization
or a fermionic local supersymmetry transformation on the worldsheet).

The restrictions (5) on the gauge parameter effectively reduce the
gauge group of superdiffeomorphisms on the worldsheet.
As a result, it is in general no longer possible to choose
the superconformal gauge, fully eliminating the functional integrals over
the worldsheet metric $\gamma^{ab}$ and the worldsheet gravitino field
$\chi^a_\alpha$ in the partition function.Instead these functional integrals
are reduced to integration over the finite number of the
conformal moduli $m_i;i=1,...,M(N)$
and the anticommuting moduli $\theta^a;a=1,...,P(N)$ of the gravitino
while the gauge-fixed superstring action is modified by the supermoduli
terms according to (1).
 The final remark about the general expression
for the correlation
function (1) concerns the dimensionalities of the  odd and even moduli spaces
as functions of the number N of the vertices.
These dimensionalities are equal to the numbers of independent
holomorphic super Beltrami differentials $\chi^a$ and $\xi^i$, dual to the
basic vectors $\theta_a(z)$ and $m_i(z)$ in the supermoduli space.
These numbers are different for the NS and Ramond sectors.
Namely, the OPE's of the stress-tensor and the supercurrent with the
unintegrated
perturbtive vertex
operators (e.g. a photon) are given by:
\eqn\grav{\eqalign{T(z)V(w)\sim{1\over{z-w}}\partial{V(w)}+O(z-w)^0\cr
G(z)V(w)\sim{1\over{z-w}}W(w)+O(z-w)^0}}
where W is some normally ordered operator of conformal dimension $1/2$.
E.g. for a picture zero photon at zero momentum: $V=c\partial{X}^m+
\gamma\psi^m$ it is easy to check that $W(w)=
{1\over2}\partial{c\psi^m}$
Now since the induced worldsheet metric and the
induced worldsheet gravitino field
can be expressed as

\eqn\grav{\eqalign{
\gamma_{ab}(z)= :\partial_a{X^m}\partial_b{X_m}:(z)+...
{\sim}T_{ab}(z)\cr
\chi_{a\alpha}(z)=:\psi^m_\alpha\partial_a{X^m}:(z)+...\sim{G_{a\alpha}}(z)
\cr
a,\alpha=1,2}}
it is clear that the $\chi_{a\alpha}(z)$,
 $\gamma_{a\alpha}(z)$ and hence the corresponding
superconformal moduli
 $\theta_a(z)$ and $m_i(z)$ behave as
$(z-z_i)^{-1};i=1,...N$ when approaching a NS vertex point.
For this reason the natural choice of the basis for
$m$ and $\theta$, consistent with their pole structures
and the holomorphic properties,
is given by
\eqn\grav{\eqalign{
\theta_a(z)=z^a\prod_{j=1}^{N}(z-z_j)^{-1}\cr
m_i(z)=z^i\prod_{j=1}^{N}(z-z_j)^{-1}\cr
a=1,...,M(N);
i=1,...,P(N)}}
The holomorphy condition requires that at the infinity
the $\theta^a$ vectors must go to zero not slower than
the supercurrent's two-point function
$lim_{z\rightarrow{\infty}}<G(0)G(z)>\sim{1\over{z^3}}$ while the $m^i(z)$
must decay as $lim_{z\rightarrow\infty}<T(0)T(z)>\sim{1\over{z^4}}$
or faster.
To see this, note that the expansions of T and G in terms of their
normal modes are
\eqn\grav{\eqalign{
T(z)=\sum_{n}{{L_n}\over{z^{n+2}}}\cr
G(z)=\sum_n{{G_n}\over{z^{n+3/2}}}}}
therefore the normal orgering of these operators at zero point
implies $n\leq{-2}$ for T and $n\leq{-}{{3}\over2}$ for G.
Under  conformal transformation $z\rightarrow{u}={1\over{z}}$
mapping the zero point to infinity, G and T transform as
\eqn\grav{\eqalign{G(z)\rightarrow{-iu^3{G(u)}}\cr
T(z)\rightarrow{u^4}T(u)+...}}
therefore the normal ordering or the regularity at the infinity requires
\eqn\grav{\eqalign{T(u)\sim{1\over{u^4}}+O({1\over{u^5}})\cr
G(u)\sim{1\over{u^3}}+O({1\over{u^4}})}}
implying the same asymptotic behaviour for $\theta_a$
and $m_i$, in the light of (7).
This condition immediately implies that
\eqn\grav{\eqalign{M(N)=N-3\cr
P(N)=N-2}}
for the scattering amplitudes in the NS-sector.
For any N unintegrated Ramond operators $V_R$
the OPE of T and V remains the same as in (6),
therefore the basis and and the number of the bosonic moduli
do not change in the Ramond sector.
However, the OPE of GSO-projected Ramond
vertex operators with the supercurrent is given by
\eqn\grav{\eqalign{G(z)V_R(w)\sim
(z-w)^{-{1\over2}}U(w)+...}}
where U(w) is some dimension 1 operator.
For this reason the supermoduli of the gravitini,
approaching the insertion points of $V_R(z_i)$,
 behave as $\chi^a(z)\sim(z-z_i)^{-{1\over2}}$.
For the scattering amplitudes on the sphere involving
the $N_1$ NS vertices
$V(z_i), i=1,...N_1$ and $N_2$ Ramond operators
$V(w_j),j=1,...,N_2$
the basis in the moduli space (involving the combinations
of quadratic and $3/2$-differentials) should be chosen as
\eqn\grav{\eqalign{\theta^a(z)=z^a
{\prod_{i=1,j=1}^{N_1,N_2}}{1\over{(z-z_i){\sqrt{z-w_j}}}}}}
The holomorphy condition (11) for these differentials
gives the total dimension of the supermoduli space
for the NSR N-point scattering amplitudes:
\eqn\grav{\eqalign{M(N)=N-3\cr
P(N)=N_1+{{N_2}\over2}-2;N_1+N_2=N}}
This concludes the explanation of the formula (1)
for the NSR scattering amplitudes.

Performing the integration over $m_i$ and $\theta_a$
in (1) using (15) we obtain

\eqn\grav{\eqalign{
<V_1^{NS-NS}(z_1,{\bar{z}}_1)...V_{N_1}^{NS-NS}(z_{N_1},{\bar{z}}_{N_1})
V_1^{RR}(w_1,{\bar{w}}_1)...
V_{N_2}^{RR}(w_{N_2},{\bar{w}}_{N_2})>
\cr
=
\int{DX}D{\psi}D{\bar\psi}D{\lbrack{ghosts}\rbrack}
e^{-S_{NSR}}
\cr
{\prod_{a=1}^{N_1+{1\over2}N_2-2}}|\delta(<\chi^a|\beta>)
<\chi^a|G_m+G_{gh}>|^2
{\prod_{i=1}^{N_1+N_2-3}}
|<\xi^i|b>\delta(<\xi^i|T>)|^2\cr
V_1^{NS-NS}(z_1,{\bar{z}}_1)...V_{N_1}^{NS-NS}(z_{N_1},{\bar{z}}_{N_1})
V_1^{RR}(w_1,{\bar{w}}_1)...V_{N_2}^{RR}(w_{N_2},{\bar{w}}_{N_2})
}}
and similarly for the open string case.
Each of
the operators $\delta(<\chi^a|\beta>)<\chi^a|G_m+G_{gh}>$
has the superconformal ghost number $+1$.
These operators  are 
the standard operators of picture-changing.
Indeed, for the particular choice of
$\chi^a=\delta(z-z_a)$ we have
\eqn\grav{\eqalign{:\Gamma:(z_a)=\delta(<\chi^a|\beta>)<\chi^a|G_m+G_{gh}>
\cr=:\delta(\beta)(G_m+G_{ghost})(z_a)=
:e^\phi(G_m+G_{ghost}):(z_a)}}
i.e. the standard expression for
the picture-changing operator and similarly for $\bar\Gamma(\bar{z_a})$
The total ghost number of the left or right picture changing insertions
following from the supermoduli integration is
therefore equal to $N_1+{1\over2}N_2-2$.
Thus if one takes all the NS-NS vertex operators, entering
 the correlator (16) or the sigma-model
terms,
at the canonical $(-1,-1)$-picture and all the RR
operators at picture $(-{1\over2},-{1\over2})$ , the integration
over the supermoduli  insures the correct
ghost number of the correlation function to cancel
the ghost number anomaly on the sphere.
Similarly, the operator
\eqn\lowen{Z=:<\xi^i|b>\delta(<\xi^i|T>)<{\bar\xi^i}|{\bar{b}}>
\delta(<{\bar\xi^i}|{\bar{T}}>):}
resulting from the integration over the
$m_i$-moduli has left and right fermionic ghost numbers $-1$.
Since we found  that the total number of these operators is
 equal to $N-3$, one has to take all the vertex operators
unintegrated
(i.e. with the left and right $+1$ $b-c$ fermionic ghost number)
so that the fermionic ghost number anomaly, equal to $-3$
on the sphere, is precisely cancelled by the moduli
integration.
The $Z$-operator (18) is a straightforward generalization
 of the picture-changing transformation (17) for the case of
the fermionic $b-c$ pictures.
In particular, the integrated and the unintegrated vertex operators
in NSR string theory  are simply two different
$b-c$ ghost picture representations of
the vertex operator. 
The Z-operator
must therefore reduce the $b-c$  ghost number of
a vertex operator by one unit,
at the same time transforming the local
operators into the non-local (i.e. the
dimension (1,1) operators integrated
 over the worldsheet). Namely,
if $c{\bar{c}}V(z,{\bar{z}})$ is an unintegrated vertex of
a conformal dimension zero, the $Z$-transformation
should give
\eqn\lowen{:Z(c{\bar{c}}V):
\sim{\int{d^2z}}V(z,{\bar{z}})+...}
i.e. the $c{\bar{c}}V$-operator is transformed into the
the worldsheet integral of $V$ plus possibly some
other terms insuring the BRST-invariance.
The non-locality of the $Z$-operator follows from the
non-locality of the delta-function of the full stress-energy tensor
in the definition (18). As previously one can
choose the conformal coordinate patches on the
Riemann surface excluding $N-3$ points corresponding to
the basis  $\xi^i(z)=\delta(z-z_{i})$, so the $Z$-operator
(18) becomes
\eqn\lowen{Z=:b{\bar{b}}\delta(T)\delta({\bar{T}}):(z,{\bar{z}})}
The expression (20) for the $Z$-operator is still not quite
 convenient for practical calculations.
Using the BRST invariance of $Z$  its OPE
properties we shall try to derive a suitable representation
for $\delta(T(z))$ (with the help of the arguments
similar to those one uses to derive the
exponential representations
for $\delta(\gamma)=e^{-\phi}$ and $\delta(\beta)=
e^\phi$)
Since the $Z$ operator must have the conformal
dimension 0, the operator $\delta(T)$ has conformal dimension
$-2$. As the full matter$+$ghost stress tensor satisfies
\eqn\lowen{
T(z)T(w)\sim{2}(z-w)^{-2}T(w)+(z-w)^{-1}\partial{T}(w)
+:TT:+...,}
the corresponding OPE for $\delta(T)$ must be given
by
\eqn\lowen{:(z-w)^2\delta(T(z))
\delta(T(w)):\sim
\delta(T(w))+...}
In addition, the $\delta(T)$-operator must satisfy
\eqn\lowen{\lbrack{Q_{brst},Z}\rbrack=:T\delta(T):=0}
since $Z$ is BRST-closed.
Moreover, as one can formally write
$:\delta(T(z)):{\sim}(T(z)-i\epsilon)^{-1}:-
:(T(z)+i\epsilon)^{-1}:$
where the
  ``inverse'' of :T(z): can be represented as
a dimension $-2$-operator
$$
T^{-1}(w)={2\over{\alpha}}\oint{{dz}\over{2i\pi}}
(w-z)^3{T(z)}$$
(where $\alpha$ is a  central charge)
so that $\lbrack{T^{-1}(z),T(z)}\rbrack=1$
we shall be looking for the representation of
$Z$ in the form
\eqn\lowen{b\delta(T(z)){\sim}:{T}^{-1}A:(z)}
where $A$ is some dimension 2 operator chosen so that $Z$ satisfies
(21),(22)
With some effort, one finds
\eqn\lowen{A(z)=b-4ce^{2\chi-2\phi}(T+b\partial{c})}
and hence
\eqn\grav{\eqalign{:b\delta(T(w)):
=\oint{dz\over{2i\pi}}(z-w)^3\lbrace{bT}-4ce^{2\chi-2\phi}
T(T-b\partial{c})\rbrace\cr
Z(w,\bar{w})=\int{d^2z}|z-w|^6
\lbrace({bT}-4ce^{2\chi-2\phi}
T(T-b\partial{c}))
({{\bar{b}{\bar{T}}}-4{\bar{c}}e^{2\bar\chi-2\bar\phi}
{\bar{T}}({\bar{T}}-{\bar{b}}\bar\partial{{\bar{c}}})})}}

By simple calculation, using
 $\lbrace{Q_{brst}},b\rbrace=T$
and $\lbrace{Q_{brst},ce^{2\chi-2\phi}}\rbrace={1\over4}-
\partial{c}ce^{2\chi-2\chi}$
it is easy to check that the expression (26) for
the fermionic ghost picture-changing operator is BRST-invariant.
The operator $Z(w,{\bar{w}})$ particularly maps the unintegrated vertices into
integrated ones.
The rules of how $Z(w,{\bar{w}})$ acts on unintegrated vertices are as follows.
Let $c{\bar{c}}V(w,{\bar{w}})$ be an unintegrated vertex operator at $w$,
where $V$ is the dimension (1,1) operator.
Writing 
\eqn\lowen{Z(w,{\bar{w}})\equiv\int{{d^2z}}|z-w|^6
R(z){\bar{R}}(\bar{z})}
where
\eqn\lowen{R(z)=bT(z)-4ce^{2\chi-2\phi}TT(z)-4bc\partial{c}
e^{2\chi-2\phi}T(z)}
is defined according to (25), one has to calculate the OPE
 between $R(z)$ and $cV(w)$ around the $z$-point.
For elementary perturbative vertices, such as a graviton,
the relevant contributions  (up to total derivatives and integrations
by parts)
will be of the order of $(z-w)^{-3}$,
cancelling the factor of $(z-w)^3$ in (25) and removing any dependence on
$w$. The  operator $W(z)$ from the $(z-w)^{-3}$
 of the OPE of the conformal dimension 1
 will then be the integrand
of the vertex in the integrated form. As for the possible
more singular terms of the OPE ,  one can show that for the perturbative
superstring vertices, such as a graviton,
 are generally the total
derivatives; higher order terms will be either total derivatives or BRST
trivial (being of the  general form $TL$ where L is BRST-closed).
The Z-operator (26) is defined for the closed string vertices.
Analogously, the operator of the $Z$-transformations
in the open string case is given by
\eqn\lowen{Z_{open}(w)=\oint_C{{dz}\over{2i\pi}}(z-w)^3R(z)}
In the latter case, the Z-transformation,
applied to the unintegrated open string vertices,
 produces the open string 
operators, integrated over some contour $C$.

As a concrete illustration, let us consider the $Z$-transformation
of the picture zero unintegrated  graviton. For simplicity
and brevity, let us consider the graviton at zero momentum.
The $k\neq{0}$ case can be treated analogously, 
even though the calculations would be a bit more cumbersome.
The expression for the unintegrated vertex operator of the graviton is
given by $$V(w){\bar{V}}({\bar{w}})\sim(c\partial{X^m}+
\gamma\psi^m)({\bar{c}}\bar\partial{X^n}+\bar\gamma\bar\psi^n)$$.
Consider the expansion of $R(z)$ with $V$ around the $z$-point
(the OPE of $\bar{R}$ with ${\bar{V}}$ can be evaluated similarly).
Let's start with the Z-transformation or the
$c\partial{X^m}$ part of $V$.
Consider the $bT$-term of $R(z)$ first.
A simple calculation gives
\eqn\grav{\eqalign{:bT:(z):c\partial{X^m}:(w)\sim(z-w)^{-2}(\partial^2{X^m}(z)+
:\partial\sigma{\partial{X^m}}:(z))+O((z-w)^{-1})}}
Next, consider the OPE of $V$ with the remaining two terms of
$R(z)$ (28).
The OPE calculation gives
\eqn\grav{\eqalign{:4ce^{2\chi-2\phi}TT-4bc\partial{c}e^{2\chi-2\phi}:(z)
c\partial{X^m}(w)\sim{4}(z-w)^{-2}:\partial\sigma{c\partial{c}}
e^{2\chi-2\phi}\partial{X^m}:(z)\cr+O((z-w)^{-1})}}
It is easy to check that
 the operator of the $(z-w)^{-2}$ term of this OPE can be represented
as
\eqn\grav{\eqalign{:\partial\sigma{c}\partial{c}e^{2\chi-2\phi}
\partial{X^m}:(z)
=-{1\over4}\partial\sigma{\partial{X^m}}(z)+\lbrace{Q_{brst}},
:ce^{2\chi-2\phi}\rbrace
\partial\sigma{\partial{X^m}:(z)}}}
Next, since $\partial\sigma(z)=-:bc:(z)$ and since
$\lbrace{Q_{brst}},c\partial{X^m}\rbrace=0$ 
due to the BRST invariance of the integrated
vertex,
we have
\eqn\lowen{\lbrack{Q_{brst}},\partial\sigma{\partial{X^m}}\rbrack=
-\lbrack{Q_{brst}},bc\partial{X^m}\rbrack=-cT\partial{X^m},}
therefore 
\eqn\grav{\eqalign{:\lbrace{Q_{brst}},ce^{2\chi-2\phi}
\rbrace\partial\sigma{\partial{X^m}}:(z)=\lbrace{Q_{brst}},
ce^{2\chi-2\chi}\partial\sigma{\partial{X^m}}\rbrace}}
since
\eqn\lowen{:ce^{2\chi-2\phi}\lbrack{Q_{brst}},\partial\sigma{\partial{X^m}}
\rbrack:
=:cce^{2\chi-2\phi}T\partial{X^m}:=0}
as $:cc:=0$.
For this reason
\eqn\lowen{4\partial\sigma{c\partial{c}}e^{2\chi-2\phi}\partial{X^m}
=-\partial\sigma{\partial{X^m}}+\lbrack{Q_{brst},...}\rbrack}
and therefore
the $(z-w)^{-2}$ order OPE term of (31) precisely
cancels the corresponding $\partial\sigma{\partial{X^m}}$-term 
of the operator product (30) of $bT$ with $c\partial{X^m}$, up to
the BRST-trivial piece.
With some more effort, one similarly can  show
that the $(z-w)^{-1}$ and other higher order
terms of the OPE of $R(z)$ with $c\partial{X^m}$ are the exact
BRST-commutators. Proceeding similarly
with the antiholomorphic OPE's, we obtain
\eqn\grav{\eqalign{:Z{c{\bar{c}}\partial{X^m}\bar\partial{X^n}}:(w,\bar{w})
=\int{d^2z}|z-w|^2\partial^2{X^m}\bar\partial^2{X^n}(z,\bar{z})
+\lbrack{Q_{brst}}...\rbrack}}
Finally, integrating twice by parts we get
\eqn\lowen{:Zc\bar{c}\partial{X^m}\bar\partial{X^n}:
=\int{d^2z}\partial{X^m}\bar\partial{X^n}(z,\bar{z})+\lbrack{Q_{brst}},...
\rbrack}
Next, consider the $\gamma\psi^m$-part of the graviton's
unintegrated vertex
(left and right-moving alike). The calculation gives
\eqn\grav{\eqalign{:R:(z)\gamma\psi^m(w)\sim
(z-w)^{-3}\partial(ce^{\chi-\phi}\psi^m)\cr+
(z-w)^{-2}\lbrack{Q_{brst},:bc\beta\psi^m:}\rbrack+
\lbrack{Q_{brst},...}\rbrack
}}
Performing the analogous calculation 
for the right-moving part and getting rid of the 
 total derivatives we obtain 
\eqn\grav{\eqalign{
:Z\gamma\bar\gamma\psi^m\bar\psi^n:
=
\lbrack{Q_{brst},...}\rbrack}}

Thus the full $Z$-transformation of the unintegrated graviton gives
\eqn\grav{\eqalign{:ZV_{grav}^{unintegr}:=
\int{d^2z}\partial{X^m}\bar\partial{X^n}(z,\bar{z})
+\lbrack{Q_{brst},...}\rbrack}}
Thus the result is given by
 the standard
integrated vertex operator of the graviton, up to BRST-trivial terms.
Note that if one naively applies the picture-changing operator $:\Gamma:$  (17)
to the integrated photon $V^{(-1)}_{ph}=
\oint{{dz}\over{2i\pi}}e^{-\phi}\psi^m$,
one gets
 $\oint{{dz}\over{2i\pi}}({\partial{X^m}+c\beta\psi^m})$.
The last term in this expression is not BRST-invariant,
therefore at the first glance the picture-changing operation seems to
violate the BRST-invariance
for some integrated vertices. This contradiction is  due to the fact that
one is not allowed to straightforwardly apply the $\Gamma$-operator
(which is the expression for the picture-changing consistent with the 
supermoduli integration) to the integrated vertices because, roughly speaking,
the $\Gamma$-operation does not  ``commute'' with the worldsheet integration.
More precisely, the contradiction arises because the
``naive'' application of the $\Gamma$ picture-changing operator
to the integrands ignores the OPE singularities
between $\Gamma$ and $Z$ (the latter can be understood as 
the ``operator of the worldsheet integration'').
On the other hand, if one uses the ``conventional'' definition of
 the picture-changing operation given by $\lbrack{Q_{brst},e^\chi{V}}\rbrack$
then one gets 
\eqn\lowen{\lbrace{Q_{brst}},{\oint{{dz}\over{2i\pi}}}
e^{\chi-\phi}\psi^m\rbrace=\oint{{dz}\over{2i\pi}}(\partial{X^m}
+\partial(ce^{\chi-\phi}\psi^m))}
The second total derivative in this expression can be thrown out
and we get the picture 0 photon.
Now it is clear that the definition of the picture-changibg
as $\lbrack{Q_{brst}},e^\chi{V}\rbrack$ is consistent with the supermoduli 
integration and the expression (17) for the local picture-changing operator,
only if one accurately accounts for the the singularities of the OPE between
$\Gamma$ and $Z$.

We have shown that the $Z$-transformation (42) of the unintegrated
graviton reproduces the full picture zero expression
for the integrated vertex operator, up to BRST-trivial terms.
Therefore the consistent procedure of the picture-changing implies
that one always applies the picture-changing operator
$\Gamma$ to unintegrated vertices, with the subsequent $Z$-transformation,
if necessary, i.e. the $\beta\gamma$ picture-changing must be followed
by the $Z$-transformation and not otherwise. 
Another useful expression is the Z-transformation of the picture-changing
operator, or the integrated form of the picture-changing.
Applying the $Z$-operator to 
$:\Gamma\bar\Gamma:=:e^{\phi+\bar\phi}G{\bar{G}}:$ one obtains
\eqn\grav{\eqalign{{(\Gamma\bar\Gamma)}_{int}(w)=:Z\Gamma:(w)
=\int{{d^2z}}|z-w|^2\lbrace:{b}\Gamma:-4:ce^{2\chi-2\phi}
(T-b\partial{c})\Gamma:\rbrace\times{c.c.}\cr
\equiv\int{d^2z}|z-w|^2{P(z)}{\bar{P}}({\bar{z}})}}

Similarly, the expression for the single left or right
integrated $\Gamma_{int}$ can be written as
\eqn\lowen{\Gamma_{int}=\oint{{dz}\over{2i\pi}}(z-w)P(z)}

The main advantage of using $:\Gamma_{int}:$ is the absence 
of singularities in the OPE between $\Gamma_{int}(w_1)\Gamma_{int}(w_2)$
which can be checked straightforwardly using the definition (44).
The singularities in the $\Gamma\Gamma$ operator products
of the usual (unintegrated) picture-changing operators are
well-known to result in complications and 
 inconsistencies in the picture-changing procedure.
As we saw, these complications and the appearance of the 
singularities are due to the fact that, strictly speaking,
 the picture-changing  procedure
 is not well-defined without the appropriate $Z$-transformations. 

\centerline{\bf 3. Ghost-Matter Mixing and Moduli Space Singularities}

The scattering amplitude (16) involves the insertion  of picture-changing
operators for bosonic and fermionic ghosts, which precise form depends
on the choice of  basis for super Beltrami quadratic and $3/2-$differentials.
It has been shown ~{\verlinde, \kn} that the scattering amplitudes
are invariant under the small variations of the Beltrami basis,
up to the total derivatives in the moduli space. In particular,
if one chooses the delta-functional basis (17), (20) for
$\xi^i$ and $\chi^a$, this symmetry  implies the independence
on  the insertion points of picture-changing operators.
The situation is more subtle, however, when the
picture-changing insertions $z_a$ of (16) coincide
with locations of the vertex operators
(which precisely is the case
for the  amplitudes involving combinations of the vertex operators
at different pictures)
The equations (6), (7) and (13) imply the singular behavior of the 
supermoduli approaching the locations of the
vertex operators. Namely, by simple conformal transformations
it is easy to check that the 
singularities of (6), (7) and (13) at $z_i$ and $w_j$ 
correspond to orbifold points
in the moduli space.
As it has been pointed out in ~{\verlinde},
if  picture-changing operators are located at the orbifold points of the
moduli space, the picture-changing gauge symmetry is reduced
to the  discrete automorphism group corresponding to all the
possible permutations of the p.c. operators between these orbifold points.
In particular, it's easy to see that for $N_1$  Neveu-Schwarz
and $N_2$ Ramond perturbative vertex operators having a 
 a total superconformal ghost number g, the volume of this
automorphism group is given by
\eqn\lowen{\Xi_{N_1,N_2}(g)=(N_1+N_2)^{N_1+{{N_2}\over2}+g}}
The appearance of this discrete group is particularly a consequence
of the polynomial property of picture-changing operators:
\eqn\lowen{:\Gamma^m::\Gamma:^n\sim:\Gamma^{m+n}:+\lbrack{Q_{BRST},...}\rbrack}
which holds as long as the picture-changing 
transformations are applied to the
perturbative string vertices, such as a graviton or a photon,
which are equivalent at  all the ghost pictures.
However, apart from the usual massless states such as a graviton or a 
photon,
the spectra of open and closed NSR strings
also contain BRST-invariant and non-trivial vertex operators which
cannot be interpreted in terms of emissions of point-like 
particles by a string. In case of an open string, an example of such a
vertex operator is an antisymmetric $5$-form, given  by
\eqn\grav{\eqalign{V_5^{open}(k)=H_{m_1...m_5}(k)
ce^{-3\phi}\psi_{m_1}...
\psi_{m_5}e^{ikX}\cr
V_{5int}^{open}(k)=H_{m_1...m_5}(k)
\oint{{dz}\over{2i\pi}}e^{-3\phi}\psi_{m_1}...\psi_{m_5}
e^{ikX}
+\lbrack{Q_{brst},...\rbrack}}}
It has been shown ~{\self} that this vertex operator
is physical, i.e. BRST-invariant and non-trivial.
The BRST non-triviality of the operator (47) requires that
the $H$ five-form is not closed:
\eqn\lowen{k_{{\lbrack}m_1}H_{m_2...m_6\rbrack}\neq{0}}
The vertex operator (47) exists only at nonzero ghost pictures
below $-3$ and above $+1$ ~{\self}, i.e.
its coupling with the ghosts is more than just an artefact of
a gauge and cannot be removed by picture-changing transformations.
This situation is referred to as the ghost-matter mixing.
In the closed string sector, an important example
of the ghost-matter mixing vertex operator can be obtained by 
multiplying the five-form (47) by antiholomorphic photonic part:
\eqn\grav{\eqalign{V_5^{(-3)}=H_{m_1...m_5m_6}(k)
c\bar{c}e^{-3\phi-\bar\phi}\psi_{m_1}...\psi_{m_5}\bar\psi_{m_6}
e^{ikX}\cr
V_{5int}^{(-3)}=H_{m_1...m_5m_6}(k)
\int{d^2z}e^{-3\phi-\bar\phi}\psi_{m_1}...
\psi_{m_5}\bar\psi_{m_6}e^{ikX}(z,\bar{z})+\lbrack{Q_{brst},...}\rbrack}}
where the 6-tensor $H_{m_1...m_6}$ is antisymmetric in the first five indices.
The BRST-invariance and non-triviality conditions for this vertex operator 
imply
\eqn\grav{\eqalign{k_{\lbrack{m_7}}H_{m_1...m_5\rbrack{m_6}}(k){\neq}0\cr
k_{m_6}H_{m_1...m_5m_6}(k)=0}}
These constraints particularly entail the gauge transformations
for the H-tensor
\eqn\lowen{H_{m_1...m_5m_6}(k)\rightarrow
H_{m_1...m_5m_6}(k)+k_{{\lbrack}m_1}R_{m_2...m_5\rbrack{m_6}}(k)}
where $R$ is a rank 5 tensor antisymmetric over the first
4 indices, satisfying $$k_{m_6}R_{m_2...m_6}=0$$
It is easy to check that the BRST constraints (50)
including the related gauge transformations (51)
eliminate 1260 out of 2520 independent components of the H-tensor.
Therefore the total number of the degrees of freedom related to the
closed string vertex operator (49) is equal to 1260. Their physical meaning 
can be understood if we note that the BRST conditions (50) imply that
for each particular polarization $m_1...m_6$ of the vertex operator (49)
the momentum $k$ must be normal to  the directions of the polarization,
i.e. confined to the four-dimensional subspace orthogonal to the
$m_1,...m_6$ directions.The number of independent polarizations of $V_5$ in
ten dimensions is equal to ${{10!}\over{4!6!}}=210$ therefore the
total number of degrees of freedom per polarization is equal to 6.
For instance consider $m_1=4,...m_6=9$ so that $k$ is polarized along 
the $0,1,2,3$
directions.Then we can make a $4+6$-split of the space-time indices
$m\rightarrow{(a,t)};a=0,...3;t=4..9;H_{m_1...m_6}\equiv{H_{t_1...t_6}}$
The tensor $H_{t_1...t_6}$ is antisymmetric in the first five indices
Since the total number of independent degrees of freedom for this polarization
is equal to 6, 
one can always choose it in the form $t_6\neq{t_i},i=1,...5$
by using suitable gauge transformations (51).
This means in turn that one can choose the basis
\eqn\grav{\eqalign{\lambda_t=H_{t_1...t_5t},\cr
t=4,...9;t\neq{t_1,...t_5}}}
with $H$ antisymmetrized over $t_1,...t_5$.
 It is easy to see that the $\lambda_t$ simply
parametrize the 6 physical degrees of freedom for this particular
polarization of $V_5$.
 To understand the meaning of $\lambda_t$-field
one has to calculate  its effective action.
 Computing the  closed string 4- point correlation function
$<V_{5}(z_1,{\bar{z_1}})...V_5(z_4,{\bar{z_4}})>$ and the three-point function
$<V_5V_5V_\varphi>$ of two $V_5$'s with the dilaton one can obtain that
they reproduce the appropriate expansion terms of the DBI effective action
for the D3-brane
\eqn\lowen{S_{eff}(\lambda)=\int{d^4x}e^{-\varphi}{\sqrt{det(\eta_{ab}+
\partial_a\lambda_t\partial_b\lambda^t)}}}
where $x$ is a Fourier transform of $k$.
The open string vertices (49) 
can also be shown to carry the Ramond-Ramond charges
which can be demonstrated by calculating
their disc correlation functions with the appropriate RR 5-form operator.
That is, the  disc correlation function $<V_5^{(-3)}(k)V_5^{(+1)}(p)
V_{RR}^{(+1/2,-1/2)}(q)>$, in which the RR vertex operator
has to be taken at the $(+1/2,-1/2)$-picture, is linear in the
momentum leading to the term $\sim{(dH)^2A_{RR}}$ in the effective 
action,  where $A$ is the Ramond-Ramond 4-form potential. This implies
 that the $dH$-field corresponds to the wavefunction
of the RR-charge carrier, i.e. of the D-brane ~{\pol}
 The BRST nontriviality 
condition (50) for the open string 
$V_5$-vertices simply means that this vawefunction does not vanish.
Thus the closed string $V_5$-operators generate the kinetic term
of the D-brane action while the open string $V_5$-vertices
account for its coupling with the RR-fields.
The fact that the closed-string amplitudes lead to the D-brane type dilaton
coupling of the effective action is related to the non-perturbative
nature of the $V_5$-vertices, which in turn is the consequence
of the ghost-matter mixing, or the picture inequivalence of $V_5$.
Let us explore this inequivalence in more details, from the point 
of view of the supermoduli geometry.
The OPE of the full matter$+$ghost supercurrent
$G(z)=-{1\over2}\psi_m\partial{X^m}-{1\over2}b\gamma+c\partial\beta
+{3\over2}\beta\partial{c}$
 with $V_5(w)$ gives
\eqn\lowen{G(z)V_5(w)\sim{-}{1\over{2(z-w)^3}}H_{m_1...m_5}
c\partial{c}e^{\chi-4\phi}
\psi_{m_1}...\psi_{m_5}e^{ikX}+...}
and  similarly for the OPE of $G$ with the left-moving part of 
the closed string $V_5$. This means that the supermoduli (8) of the
 gravitini behave as $\theta^a(z)\sim(z-z_k)^{-3}$ as they approach
insertion points of $V_5$.
Such a behavior of the supermoduli approaching the 
$V_5$ operators is much more singular than of those approaching
the usual perturbative  vertices (8).
These singularities no longer correspond to the orbifold points
of the moduli space. Instead they entirely overhaul the  moduli space topology,
effectively creating boundaries and global curvature singularities.
To illustrate this consider the worldsheet metric
\eqn\lowen{ds^2=dzd\bar{z}+z^{-\alpha}(dz)^2+{\bar{z}}^{-\alpha}(d\bar{z})^2}
In terms of  the $r,\varphi$ coordinates where
$z=re^{i\varphi},{\bar{z}}=re^{-i\varphi}$ this metric is
given by
\eqn\lowen{(1+2r^{-\alpha}cos(\alpha-2)\varphi)dr^2+
(1-2r^{-\alpha}cos(\alpha-2)\varphi)r^2d\varphi^2-4r^{1-\alpha}
sin(2-\alpha)\varphi{dr}d\varphi}
The area of the disc of radius $\epsilon$ surrounding
the origin point is given by
\eqn\lowen{A=\int_0^\epsilon{dr}\int_0^{2\pi}{d\varphi}{\sqrt{\gamma}}
=\int_0^{2\pi}d\varphi{\int_0^\epsilon}d{r}r{\sqrt{1-2r^{-2\alpha}}}}
This integral is of the order of $\epsilon^{2}$ for positive $\alpha$,
$\epsilon^{2-\alpha}$ for $0\leq\alpha<2$
(this value reflects the deficit of the angle near the orbifold points)
but it diverges if $\alpha\geq{2}$ which means that for
$\alpha=3$ the disc is no longer compact but the origin point is blown up
to become a global singularity.
For this reason the scattering amplitudes in the presence of
the $V_5$-vertices
are no longer invariant under the discrete automorphism group (45).
This means that the OPE's involving the $V_5$-operators become
picture-dependent. Therefore, in order to correctly describe the  
physical processes involving the $V_5$-operators one has to sum over
 all the previously  equivalent gauges, i.e. all the
admissible pictures (corresponding to the possible locations
of the picture-changing operators (17) at the singularity points
of the moduli space) and normalize
 by the volume (45) of the original gauge group of automorphisms.
The presence of the $V_5$-operators also modifies the suitable choice of the 
basis for the supermoduli and the number of independent
$3/2$-differentials
 For the amplitudes involving
the total number $N_1$  of the $V_5$-insertions
and $N_2$ of the standard perturbative vertices (without 
the ghost-matter mixing) the basis for $\theta^a$ is given by
\eqn\lowen{\theta_a(z)=z^a\prod_{i=1}^{N_1}(z-z_i)^{-3}\prod_{j=1}^{N_2}
(z-w_j)^{-1}}
Accordingly the holomorphy condition (11) implies that in this case
the number of the $3/2$-differentials (equal to  the number
of  the picture-changing insertions) is equal to
\eqn\lowen{P(N)=3N_1+N_2-2}.

As previously, the supermoduli integration insures the correct total 
ghost number
of the vertex operators in the amplitude provided that
the perturbative vertices of the generating sigma-model 
are all taken at pictures
$-1$ while the ghost-matter mixing $V_5$-vertices are at the picture
$-3$.

When a string propagating in flat space-time emits  particles or  solitons,
the background is perturbed by the appropriate vertex operators.
These operators multiplied by the corresponding space-time fields
must be added to the original NSR superstring action (2).
The condition of conformal invariance then leads to the
effective equations of motion for these space-time fields and
to the corresponding effective field theory.
We conclude this section by writing down the generating
functional for the NSR string sigma-model, free of the 
picture-changing ambiguities.
The partition function of  superstring theory perturbed by
the set $\lbrace{V_i}\rbrace$ of physical vertex operators is given by
\eqn\grav{\eqalign{Z(\varphi_i)=
\int{DXD}\psi{D{\bar\psi}}{D}\lbrack{ghosts}\rbrack
{e^{-S_{NSR}+\varphi_i{V^i}}}\rho_{(\Gamma;Z)}\rho_{(\bar\Gamma;{\bar{Z}})}}}
where $\rho_{(\Gamma;Z)}$  is the picture-changing
factor due to the appearance of the supermoduli when one expands in 
$\varphi_i$. Though this factor enters differently for  
each term of the expansion (as it is clear from (16)), 
it is straightforward to check that, due to the ghost number conservation,
one can recast it in the invariant form (independent on the order of expansion)
\eqn\grav{\eqalign{\rho_{(\Gamma;Z)}=
\sum_{m,n=0}^{\infty}\Xi_{\xi}^{-1}(n)\Xi_\chi^{-1}(m)
\sum_{\lbrace{\xi^{(1)},..\xi^{(n)},
\chi^{(1)}...\chi^{(m)}}\rbrace}\delta(<\chi^{(1)}|\beta>)
<\chi^{(1)}|G>...\cr
\delta(<\chi^{(m)}|\beta>)<\chi^{(m)}|G>
<\xi^{(1)}|b>\delta(<\xi^{(1)}|T>)...<\xi^{(n)}|b>\delta(<\xi^{(n)}|T>)}}
and accordingly for $\rho_{({\bar\Gamma};{\bar{Z}})}$
where the sum over $\xi^{(i)}$ and $\chi^{(i)}$ implies the summation over all
the basic vectors of the $(m,n)$-dimensional
spaces of super Beltrami differentials.
Note that due to the ghost number anomaly cancellation condition,
 for any N-point correlator appearing as a result of expansion in
in $\lambda$ only the $m=N-3$ and $n={N_1}+{1\over2}{N_2}-2$
terms of $\rho_{(\Gamma;Z)}$ contribute to the sigma-model.
$\Xi_\xi(m)$ and $\Xi_\chi(n)$ are the volumes of the
symmetry groups related to the picture-changing gauge symmetry.
(defined separately for the left and the right picture-changing).
In the important case when the basic vectors
are chosen at the orbifold points of the moduli spaces,  the
volumes are given by the relation (45). It's easy to see
that in the picture-independent case inserting the
$\rho_{(\Gamma,Z)}$-factor in the partition function can be reduced to the
trivial statement that if one  sums over N equivalent amplitudes with 
ghost picture combinations of the vertices and then divides
by N, one gets the value of the amplitude. In the ghost-matter mixing
cases involving the global singularities in the moduli space, the situation
is more complicated. Thus the generating functional (60), (61)
 of the NSR sigma-model
is a straightforward consequence of the expression (16) for the scattering 
amplitudes derived from the supermoduli integration. As was already
said above, the operators $V_i$ of the sigma-model action (60) 
are taken unintegrated at picture $-1$
(and those with the ghost-matter mixing are at the picture $-3$). These 
operators are generally BRST invariant 
but not invariant under superdiffeomorphisms
which simply means that
the gauge symmetry, related to global conformal 
transformations,
is fixed from the very beginning in the model (60),(61).
The choice of the insertion points of the vertex operators corresponds
to the choice of the Koba-Nielsen's measure in the correlation functions.
Indeed, the $Z$-operators appearing as a result of the integration over the
 bosonic moduli, transform $N-3$ out of $N$ vertex operators into
the integrated ones, while the remaining 3 are left in the 
unintegrated form $\sim{c\bar{c}V({z_i,{\bar{z_i}}});i=1,2,3.}$. 
Then the $c$ and ${\bar{c}}$-fields
contribute the factor of $\prod_{i,j}|z_i-z_j|^2$ which precisely
is the invariant Koba-Nielsen's measure ~{\koba}
 necessary for the calculations of the string scattering amplitudes.

\centerline{\bf Conclusions}

In this paper we have constructed the sigma-model for NSR superstrings,
leading to the consistent string perturbation theory, free of the
picture-changing ambiguities.
The important element of the  construction is the appearance of the 
$b-c$-picture changing operators defined by (18),(20).
The $b-c$ picture-changing operators particularly map
unintegrated vertices into integrated ones, up
to total derivatives and BRST-trivial terms.
In case if  the $Z$-operator (18),(20) acts on the integrated vertices,
one obtains the ``double-integrated'' representations
of the vertex operators, etc.
Just like not all the physical vertex operators
can be represented at picture zero (due to the global singularities 
in the space of supermoduli), the global singularities in the spaces
of the bosonic moduli lead to the $b-c$ picture inequivalence
for some vertex operators.As a result there are the physical operators
existing at some particular $b-c$ pictures with no 
equivalent version at other pictures, e.g. the operators 
that can be represented in the
integrated pictures, but not in the unintegrated $cV$-form.
In addition, the OPE of the $Z(w,\bar{w})$ operator
defined  by (18),(20), with unintegrated perturbative massless vertices
(such as graviton) $c{\bar{c}}W(w,\bar{w})$ 
gives an expression independent on $w$ (i.e. the integral of
$W$ over the worldsheet). In general, however, it is
possible that the global moduli space singularities
may lead to the appearance of the terms of the form
${\sim}W(z)\sim\int{d^2w}|z-w|^{2n}U(w,\bar{w})$ 
in the expressions for physical
vertex operators, where $U$ is some operator of conformal dimension
$(n+1,n+1)$.
The important example of the $b-c$ picture inequivalence is the
5-form (49) at the picture $+1$.
The naive picture $+1$-expression for the five-form (47) 
$\sim{H_{m_1...m_5}\oint{e^\phi{\psi_{m_1}...\psi_{m_5}}}}e^{ikX}$
is not BRST-invariant due to the non-commutation with the supercurrent part 
of $Q_{brst}$. To insure the BRST-invariance of the five-form (47)
at the $+1$-picture one has to add the double-integrated $b-c$ ghost terms
so that the full BRST-invariant expression for the $V_5^{(+1)}$
is given by
\eqn\grav{\eqalign{V_5^{(+1)}=H_{m_1...m_5}(k)\oint{{dz}\over{2i\pi}}
{\lbrace}e^\phi\psi_{m_1}...\psi_{m_5}e^{ikX}\cr
+2{\lbrack}{{\hat{b}}_3}c\partial{c}
{e^\chi}\partial\chi(\psi_{m_1}...\psi_{m_5}(\psi_m\partial{X^m})
+\psi_{{\lbrack}m_1}...\psi_{m_4}(\partial{X_{m_5\rbrack}}
(\partial\phi-\partial\chi)+\partial^2{X_{m_5\rbrack}})\cr
+i\psi_{m_1}...\psi_{m_5}((k\psi)(\partial\phi-\partial\chi)+(k\partial\psi))
{\rbrack}e^{ikX})\cr
+{1\over{40}}{\lbrack}({\hat{T}}^\chi)_6(\partial\phi-\partial\chi)
c\partial{b}be^{2\phi-\chi}\psi_{m_1}...\psi_{m_5}(\psi\partial^2{X})
{\rbrack}e^{ikX}\rbrack\rbrace\cr
k_{{\lbrack}m_1}H_{m_2...m_6\rbrack}\neq{0}}}
where the operators with the hat acting on an arbitrary operator
$A(z)$ are defined as
\eqn\grav{\eqalign{
{\hat{b}}_nA(z)=\oint{{dw}\over{2i\pi}}(w-z)^{n+1}:b(w)A(w):\cr
({\hat{T}}^\chi)_nA(z)=\oint{{dw}\over{2i\pi}}(w-z)^{n+1}:T^\chi(z)A(z):\cr
T^\chi(z)=\partial\chi\partial\chi(z)-\partial^2\chi(z)}}
It's easy to show that the OPE of the full stress-tensor with $V_5^{(+1)}$
is given by $T(z)V_5^{(+1)}(w)\sim(z-w)^{-3}U(w)+...$
while the OPE of T with usual integrated vertices contains no
singularities at all (the last OPE can be represented as a worldsheet integral
of a total derivative). 
At the same time, the OPE of the supercurrent with the
$V_5^{(+1)}$-operator contains only the regular
$(z-w)^{-1}$ orbifold-type singularity (unlike the case of the
five-form at the picture $-3$).
Therefore the bosonic moduli
behave as $(z-z_i)^{-3}$ approaching the $V_5^{(+1)}$ insertion points.
Hence the $V_5^{(+1)}$-insertions correspond to the global
singularities in the spaces of the bosonic moduli,
while the anticommuting moduli have the usual orbifold singularities
at these insertion points.
Thus the picture $+1$ and $-3$ five-form operators (47) and (62) , while 
representing the same massless physical state, are ontologically
different from the point of view of the moduli space geometry:
the first originates from the global singularities
of the bosonic moduli space, 
while the second corresponds to those in the spaces
of the anticommuting supermoduli (accordingly, they are related
to the orbifold points in the fermionic and bosonic moduli 
spaces). As has been noted in this paper,
the global singularities in the superconformal moduli spaces
lead to the breaking of the picture-changing gauge symmetry.
This results in the picture-dependence of the operator products
involving the ghost-matter mixing vertex operators (related
to the creation of the D brane-like solitons (49),(52),(53)).
As a result, due to this picture-dependence , the conformal $\beta$-function 
equations involving these vertex operators are entirely different from the 
usual ones; namely, these equations become stochastic, having the form of the
non-Markovian Langevin equations, where the operator of the stochastic 
noise is given by the worldsheet integral of $V_5$, cut at a certain
scale $\Lambda$. The worldsheet cutoff $\Lambda$ corresponds to the 
stochastic time of the process, and the memory structuer of the noise
is determined by the worldsheet correlators
of $V_5$. In our next paper, currently in progress and close to the conclusion,
we will show how the equations of the hydrodynamics 
(Navier-Stokes and the passive scalar equations) emerge
for the dilaton and the D-brane's U(1) field in the ghost-matter mixing
 backgrounds. These stochastic processes possibly play a vital role
in the gauge-string correspondence ~{\malda, \ampf, \witten},
in the formation of  the space-time geometry of our world,
and may hint at the existence of
 the deep connections between string theory
and the physics of hydrodynamics, chaos and turbulence.
We will discuss these connections in details in
our next paper to come.

\centerline{\bf Acknowledgements}

The author is grateful to A. Chamseddine, I. Giannakis,
A.M. Polyakov and W. Sabra for useful comments and discussions.

\listrefs
\end